\documentclass{article}
\usepackage[centertags]{amsmath}
\DeclareMathOperator{\SN}{sn} \DeclareMathOperator{\CN}{cn}

\begin{document}

\title{Representations and classification of traveling wave solutions to Sinh-G{\"o}rdon equation}

\author{ Cheng-shi Liu \\Department of Mathematics\\Daqing Petroleum Institute\\Daqing 163318, China
\\Email: chengshiliu-68@126.com}

 \maketitle

\begin{abstract}
Two concepts named atom solution and combinatory solution are
defined. The classification of all single traveling wave atom
solutions to Sinh-G{\"o}rdon equation is obtained, and qualitative
properties of solutions are discussed. In particular, we point out
that some qualitative properties derived intuitively from dynamic
system method aren't true. In final, we prove that our solutions to
Sinh-G{\"o}rdon equation include all solutions obtained in the
paper[Fu Z T et al, Commu. in Theor. Phys.(Beijing) 2006 45 55].
Through an example, we show how to give some new identities on
Jacobian elliptic functions.


\end{abstract}

\section{Introduction}
The construction of traveling wave solutions for nonlinear evolution
equations is one of important aims in nonlinear science. Various
methods are proposed for this purpose (for example,see[1-28,30-35]).
By these methods, a large number of nonlinear equations are solved
and their abundant traveling wave solutions are obtained. Although
these method give some applicable results, some problems are also
need further studied. For example, can classification of all single
traveling wave solutions be given? How to construct new solutions?
How to decide that a solution is novel?  In order to solve these
problem, we first give the following definitions.

\textbf{Definition 1}: For a given nonlinear evolution, if its one
traveling wave solution is obtained by direct integral method, we
call this solution an atom solution.

\textbf{Definition 2}: A solution constructed piecewise in terms of
atom solutions is called a combinatory solution.

For these equations which can be directly reduced to an elementary
integral form, we can obtain  classifications of their all single
traveling wave atom solutions. Based on these atom solutions, by
piecewise construction, we can give some combinatory solutions with
some new properties, such as peakon solutions[27], compacton
solutions[28], etc. Therefore, to give classification of all atom
solutions is an important aim. In my early papers\cite{L3,li4}, I
have obtained classifications of atom solutions for some nonlinear
equations, in which I don't define the concept of atom solution.
Hence, here I must point out that it is just for atom solutions to
give those corresponding classifications. In present paper, I
discuss Sinh-G{\"o}rdon equation. It reads
\begin{equation}
u_{xt}=\alpha\sinh u,
\end{equation}
which is applied to integrable quantum field, noncommutative field
theories, fluid dynamics, geometry {\cite{chern}}, etc.  Its some
exact solutions have been obtained by some authors
{\cite{fu,pa,si,ua,xie,wa}}. In order to solve its
 new single traveling wave solutions, Fu \emph{et al} \cite{fu} try to
 introduce two variable transformations, and  obtain some special form solutions under
some special cases. But we must point out that it is a rather easy
thing to obtain all single traveling wave atom solutions to
Sinh-G{\"o}rdon equation. Using elementary integral method and
complete discrimination system for polynomial[5-8], we obtain this
complete result. On the other hand, from qualitative analysis,
intuitively, no periodic solution exists for Sinh-G{\"o}rdon
equation. But we have  several periodic solutions indeed.  Hence it
seems that there is a contradictory. We discuss this problem and
show that sometimes qualitative conclusion may be not true.

Some authors don't believe that these solutions obtained by our
method are all atom solutions of Sinh-G{\"o}rdon equation. They
point out to me that professors Liu Shi-Kuo \emph{et al}
\cite{l1,l2} have given a lot of different solutions to elliptic
equation, and hence ask me how I can transfer their solutions into
our solutions. In the present paper, I prove this conclusion. In
other words, their solutions only give some new representations of
solutions to elliptic equation rather than new solution. In
particular, I must emphasize that my proofs are not direct.
Furthermore, from these proofs, we can obtain some interesting new
identities on Jacobian elliptic functions. Through an example, we
show how to do this thing.

This paper is organized as follows: In Section 2, we give the
classification of all traveling wave atom solutions to
Sinh-G{\"o}rdon equation. In Section 3, we discuss qualitative
analysis of solutions for Sinh-G{\"o}rdon equation. In Section 4, we
prove that all solutions to Sinh-G{\"o}rdon equation given in
reference \cite{fu} can be represented by our solutions. In
particular, through an example, we show how to give some new
identities on Jacobian elliptic functions. The last section is a
short summary.

\section{Classification of traveling wave atom solutions to Sinh-G{\"o}rdon equation}
We take traveling  wave transformation
\begin{equation}
u=u(\xi), \ \ \xi=kx+\omega t.
\end{equation}
Substituting Eq.(2) into Eq.(1) yields
\begin{equation}
k\omega u''=\alpha\sinh u.
\end{equation}
By integrating  Eq.(2) once and rewriting it as integral form, we
have
\begin{equation}
\pm(\xi-\xi_0)=\int\frac{\mathrm{d}u}{\sqrt{\frac{2\alpha}{k\omega}(\cosh
u+c)}}.
\end{equation}
It is rather natural to take variable transformation as follows
\begin{equation}
w=\exp u.
\end{equation}
Then $w>0$ and Eq.(4) becomes
\begin{equation}
 \pm(\xi-\xi_0)=\int\frac{1}{\sqrt{\frac{\alpha}{k\omega}w(w^2+2cw+1)}}\mathrm{d}w.
\end{equation}
Denote $ \triangle=4c^2-4 $. If $\frac{\alpha}{k\omega}>0,$
there are the following four cases to be discussed.\\

Case 1: $ \triangle=0 $, that is,
 $ c=\pm1$ . Since $ w>0 $, if
$\frac{\alpha}{k\omega}<0,$ then no solutions to  Eq.(6). When
$\frac{\alpha}{k\omega}>0,$ there are the following two cases

If  $c=-1$, we have

\begin{equation}
u=\ln\tanh^2(\frac{1}{2}\sqrt{\frac{\alpha}{k\omega}}(\xi-\xi_0)),
\end{equation}
\begin{equation}
u=\ln\coth^2(\frac{1}{2}\sqrt{\frac{\alpha}{k\omega}}(\xi-\xi_0)).
\end{equation}\\
If $c=1$, we have
\begin{equation}
u=\ln\tan^2(\frac{1}{2}\sqrt{\frac{\alpha}{k\omega}}(\xi-\xi_0)).
\end{equation}\\

Case 2: $ \triangle>0 $.  Suppose that $ \rho<\beta<\gamma $, one of
them is zero, and others two roots of $ w^2+2cw+1=0 $. For
$\frac{\alpha}{k\omega}>0,$ when $\rho<w<\beta $ and $w>0$, we have
\begin{equation}
u=\pm\ln\{\rho+(\beta-\rho)\SN^2(\frac{\sqrt{\gamma-\rho}}{2}\sqrt{\frac{\alpha}{k\omega}}(\xi-{\xi}_0)),m)\}.
\end{equation}
Concretely, the solutions are given by
\begin{equation}
u=\pm\ln\{m\SN^2(\frac{1}{2\sqrt{m}}\sqrt{\frac{\alpha}{k\omega}}(\xi-{\xi}_0)),m)\}.
\end{equation}
where $0<m<1$.

 When $ w>\gamma $ and $w>0$, we have
\begin{equation}
u=\pm\ln\{\frac{\gamma-\beta\SN^2(\frac{\sqrt{\gamma-\rho}}{2}\sqrt{\frac{\alpha}{k\omega}}
(\xi-{\xi}_0)),m)}{\CN^2(\frac{\sqrt{\gamma-\rho}}{2}\sqrt{\frac{\alpha}{k\omega}}(\xi-{\xi}_0)),m)}\},
\end{equation}
where $ m^2=\frac{\beta-\rho}{\gamma-\rho} $. Concretely, the
solutions are given by
\begin{equation}
u=\pm\ln\{\frac{1-m^2\SN^2(\frac{1}{2\sqrt{m}}\sqrt{\frac{\alpha}{k\omega}}
(\xi-{\xi}_0)),m)}{m\CN^2(\frac{1}{2\sqrt{m}}\sqrt{\frac{\alpha}{k\omega}}(\xi-{\xi}_0)),m)}\},(c<-1),
\end{equation}
and
\begin{equation}
u=\pm\ln\{\frac{m\SN^2(\frac{1}{2\sqrt{m}}\sqrt{\frac{\alpha}{k\omega}}
(\xi-{\xi}_0)),m)}{\CN^2(\frac{1}{2\sqrt{m}}\sqrt{\frac{\alpha}{k\omega}}(\xi-{\xi}_0)),m)}\},(c>1).
\end{equation}

For $\frac{\alpha}{k\omega}<0,$  when $ \beta<w<\gamma $ and $w>0$,
we have
\begin{equation}
u=\pm\ln\{\beta-(\beta-\gamma)\SN^2(\frac{\sqrt{\gamma-\rho}}{2}\sqrt{-\frac{\alpha}{k\omega}}(\xi-{\xi}_0)),m)\},
\end{equation}
where $ m^2=\frac{\beta-\rho}{\gamma-\rho} $. Concretely, we have
\begin{equation}
u=\pm\ln\{\frac{1-(1-m^2)\SN^2(\frac{1}{2\sqrt{m}}\sqrt{-\frac{\alpha}{k\omega}}
(\xi-{\xi}_0)),m)}{m\CN^2(\frac{1}{2\sqrt{m}}\sqrt{-\frac{\alpha}{k\omega}}(\xi-{\xi}_0)),m)}\},(c<-1).
\end{equation}
\textbf{Remark 1}: Because two roots of the equation $w^2+2cw+1=0$
have the same signs,  we have $\rho=0$ or $\gamma=0$. Moreover, the
product of two nonzero roots is 1. These facts are important for
verifying the above solutions.\\

Case3: $\triangle<0 $. When $\frac{\alpha}{k\omega}<0,$  no
solutions to Eq.(6). When $\frac{\alpha}{k\omega}>0,$ we have
\begin{equation}
u=\pm\ln\{\frac{2}{1+\CN(\sqrt{\frac{\alpha}{k\omega}}(\xi-{\xi}_0)),m)}-1\},
\end{equation}\\
where $ m^2=\frac{1}{2}(1-c) $.\\

Expressions (7-9),(11)(13)(14)(16)(17) are all solutions to Eq.(4).
Thus we give classification of all solutions to Eq.(3). These
solutions are all possible  single traveling wave atom solutions to
Sinh-G{\"o}rdon equation.
This is a complete result.\\

\textbf{Remark 2}: It is easy to see that if $u$ is a solution of
Sinh-G{\"o}rdon equation, then $-u$ is also a solution. For example,
$u=\ln\tan^2(\frac{1}{2}\sqrt{\frac{\alpha}{k\omega}}(\xi-\xi_0))$
is a solution, then
$u=\ln\cot^2(\frac{1}{2}\sqrt{\frac{\alpha}{k\omega}}(\xi-\xi_0))$
is also a solution. But we can transfer these two solutions each
other by taking $\xi_0$ as $\xi_0+\frac{\pi}{2}$. For convenience,
we add a sign $\pm$ in front of the expressions (10)-(17).

\section{Qualitative analysis}
Let us analysis qualitative properties of the above traveling wave
solutions of Sinh-G{\"o}rdon equation. Here we use dynamic system
method. Rewrite sinh-G{\"o}rdon equation as
\begin{eqnarray}
u'=v,\\
v'=\frac{\alpha}{k\omega}\sinh u.
\end{eqnarray}
Its unique singular point is $(0,0)$, in which eigenvalue of
Jacobian matrix satisfies $\lambda^2=\frac{\alpha}{k\omega}$.
Therefore, for $\frac{\alpha}{k\omega}>0$, (0,0) is a saddle point,
and for $\frac{\alpha}{k\omega}<0$, (0,0) is a center point.
Furthermore, according to the theory of dynamic system, it seems
that it is impossible to give periodic solution to sinh-G{\"o}rdon
equation for $\frac{\alpha}{k\omega}>0$. But in fact solution (9) is
a periodic solution, and solutions (13) and (14) are the double
periodic solutions. This is a contradictory. Why? How to solve this
problem?

It is easy to prove that expressions (9), (13) and (14) are indeed
the solutions of Sinh-G{\"o}rdon equation through substituting
directly these solution into it. So we have no doubt for these
solutions. If we notice that solution (9) is discontinuous periodic
function, then we can conclude that there isn't contradictory at all
because that so-called periodic solution needed for saddle point
must be continuous function. Based on the same reason, solutions
(13), (14) and (17) are also reasonable.

In fact, for example, the solution (9) has period $\pi$ and
discontinuous points $k\pi/2$ with $k$ arbitrary integer, and it is
unbounded and monotonic in $(-\pi/2+k\pi, k\pi)$ or $(k\pi,
k\pi+\pi/2)$, etc. If starting point is given, then the trajectory
will move continuously along a branch in a periodic interval, and
can't move from one periodic interval to another periodic interval.
Thus these so-called periodic solutions are not realizable periodic
solutions. We give this kind of solution a name by

\textbf{Definition 3}: Nearby a saddle point, if a solution is a
discontinuous periodic solution, then we call it a formal-periodic
solution.

Therefore, solutions (9),(13),(14),(16) and (17) are formal-periodic
solutions to Sinh-G{\"o}rdon equation.

A problem need studied further is when a formal-periodic solution
exists.

\section{Representations of solutions }

In Ref.{\cite{fu}}, Fu \emph{et al} introduce two transformations
\begin{equation}
u=2\sinh^{-1}v
\end{equation}
and
\begin{equation}
u=2\cosh^{-1}v
\end{equation}
to try to obtain new solutions to Sinh-G{\"o}rdon equation. But if
we write these two transformations as more explicit forms
\begin{equation}
 u=2\ln(v+\sqrt{v^2+1})
 \end{equation}
and
\begin{equation}
u=2\ln(v\pm\sqrt{v^2-1}),
\end{equation}
we can see that these transformations are rather complicated.
However, for example, if we take
\begin{equation}
w=(v+\sqrt{v^2+1})^2,
\end{equation}
 then we have
\begin{equation}
 v=\frac{w-1}{2\sqrt w},
\end{equation}
 and hence we can transfer
those results in Ref.\cite{fu} into our expressions. According to
our method used here, we can prove that all solutions in
Ref.\cite{fu} are the special cases of our solutions. For shortly,
we only consider the first kind transformation, i.e.,
$u=2\sinh^{-1}v$, and the first solutions $u_1$ in Ref.\cite{fu}.
The second kind transformation and other solutions can be dealt with
similarly. In order to illustrate my method, as an example, we solve
the solutions of $(u')^2=1-u^2$. Obviously,  its solutions are
\begin{equation}
u(\xi)=\pm\sin(\xi-\xi_0).
\end{equation}
On the other hand, if we take  transformation of variable
\begin{equation}
u=\frac{1-w^2}{1+w^2},
\end{equation}
then corresponding integral is
 \begin{equation}
\pm(\xi-\xi_1)=-2\int \frac{\mathrm{d}w}{1+w^2}=-2\arctan w.
 \end{equation}
 Hence the solutions are given by
 \begin{equation}
w=\pm\tan(\frac{1}{2}(\xi-\xi_1)),
 \end{equation}
 that is
 \begin{equation}
u(\xi)=\frac{1-\tan^2(\frac{1}{2}(\xi-\xi_1))}{1+\tan^2(\frac{1}{2}(\xi-\xi_1))}
=\cos(\xi-\xi_1).
 \end{equation}
 If we take $\xi_1=\xi_0+\frac{\pi}{2}$, then
 \begin{equation}
\sin(\xi-\xi_0)=\cos(\xi-\xi_1);
 \end{equation}
 If we take $\xi_1=\xi_0+\frac{3\pi}{2}$, then
\begin{equation}
-\sin(\xi-\xi_0)=\cos(\xi-\xi_1).
 \end{equation}
 This show that these two  solutions in distinct forms are indeed the same solution with
 distinct integral constants.

 Now we consider the following elliptic equation
 \begin{equation}
(v')^2=(v^2+1)(v^2-c),
 \end{equation}
 where we take $c>0$. We use two transformations to solve the corresponding integral so
that we obtain
 a solution which has two distinct forms. Firstly, by taking the
transformation
 \begin{equation}
v=\pm\sqrt y,
 \end{equation}
 we have
 \begin{equation}
\pm2(\xi-\xi_0)=\int \frac{\mathrm{d}y}{\sqrt{y(y+1)(y-c)}}.
 \end{equation}
 When $y>c$, the corresponding solution is
 \begin{equation}
y=\frac{c}{\CN^2(\sqrt{c+1}(\xi-\xi_0),m_1)},
 \end{equation}
 where $m_1^2=\frac{1}{1+c}$. The corresponding $v$ is
 \begin{equation}
v=\pm\frac{\sqrt c}{\CN(\sqrt{c+1}(\xi-\xi_0),m_1)}.
 \end{equation}

 Secondly, by taking the transformation
 \begin{equation}
v=\frac{w-1}{2\sqrt w},
 \end{equation}
where $w>0$, integral becomes
 \begin{equation}
\pm(\xi-\xi_1)=\int \frac{\mathrm{d}w}{\sqrt{w(w^2-2(2c+1)w+1)}}.
 \end{equation}
Since $v^2>c$,  we have $w>2c+1+2\sqrt{c(c+1)}$, and hence the
corresponding solution $w$ is given by
\begin{equation}
w=\frac{2c+1+2\sqrt{c(c+1)}-(2c+1-2\sqrt{c(c+1)})\SN^2(\frac{\sqrt{2c+1+2\sqrt{c(c+1)}}}{2}(\xi-\xi_1),
m_2) } {\CN^2(\frac{\sqrt{2c+1+2\sqrt{c(c+1)}}}{2}(\xi-\xi_1),
m_2)},
\end{equation}
where $m_2^2=\frac{2c+1-2\sqrt{c(c+1)}}{2c+1+2\sqrt{c(c+1)}}$.
Therefore, we have
\begin{equation}
v=\frac{\frac{2c+1+2\sqrt{c(c+1)}-(2c+1-2\sqrt{c(c+1)})\SN^2(\frac{\sqrt{2c+1+2\sqrt{c(c+1)}}}{2}(\xi-\xi_1),
m_2) } {\CN^2(\frac{\sqrt{2c+1+2\sqrt{c(c+1)}}}{2}(\xi-\xi_1),
m_2)}-1}{2\sqrt{\frac{2c+1+2\sqrt{c(c+1)}-(2c+1-2\sqrt{c(c+1)})\SN^2(\frac{\sqrt{2c+1+2\sqrt{c(c+1)}}}{2}(\xi-\xi_1),
m_2) } {\CN^2(\frac{\sqrt{2c+1+2\sqrt{c(c+1)}}}{2}(\xi-\xi_1),
m_2)}}}.
\end{equation}
When $\xi=\xi_1$, we have
\begin{equation}
w(\xi_1)=2c+1+2\sqrt{c(c+1)},
\end{equation}
that is
\begin{equation}
v(\xi_1)=\frac{2c+2\sqrt{c(c+1)}}{2\sqrt{2c+1+2\sqrt{c(c+1)}}},
\end{equation}
and hence we have
\begin{equation}
\pm\frac{\sqrt
c}{\CN(\sqrt{c+1}(\xi_1-\xi_0),m_1)}=\frac{2c+2\sqrt{c(c+1)}}{2\sqrt{2c+1+2\sqrt{c(c+1)}}}.
\end{equation}
Thus we conclude that if
\begin{equation}
\CN(\sqrt{c+1}(\xi_1-\xi_0),m_1)=\frac{\sqrt{2c+1+2\sqrt{c(c+1}}}{1+\sqrt{c+1)}},
\end{equation}
then
\begin{eqnarray}
v=\frac{\sqrt c}{\CN(\sqrt{c+1}(\xi-\xi_0),m_1)}\cr
=\frac{\frac{2c+1+2\sqrt{c(c+1)}-(2c+1-2\sqrt{c(c+1)})\SN^2(\frac{\sqrt{2c+1+2\sqrt{c(c+1)}}}{2}(\xi-\xi_1),
m_2) } {\CN^2(\frac{\sqrt{2c+1+2\sqrt{c(c+1)}}}{2}(\xi-\xi_1),
m_2)}-1}{2\sqrt{\frac{2c+1+2\sqrt{c(c+1)}-(2c+1-2\sqrt{c(c+1)})\SN^2(\frac{\sqrt{2c+1+2\sqrt{c(c+1)}}}{2}(\xi-\xi_1),
m_2) } {\CN^2(\frac{\sqrt{2c+1+2\sqrt{c(c+1)}}}{2}(\xi-\xi_1),
m_2)}}};
 \end{eqnarray}
if
\begin{equation}
\CN(\sqrt{c+1}(\xi_1-\xi_0),m_1)=-\frac{\sqrt{2c+1+2\sqrt{c(c+1}}}{1+\sqrt{c+1)}},
\end{equation}
then
\begin{eqnarray}
v=-\frac{\sqrt c}{\CN(\sqrt{c+1}(\xi-\xi_0),m_1)}\cr
=\frac{\frac{2c+1+2\sqrt{c(c+1)}-(2c+1-2\sqrt{c(c+1)})\SN^2(\frac{\sqrt{2c+1+2\sqrt{c(c+1)}}}{2}(\xi-\xi_1),
m_2) } {\CN^2(\frac{\sqrt{2c+1+2\sqrt{c(c+1)}}}{2}(\xi-\xi_1),
m_2)}-1}{2\sqrt{\frac{2c+1+2\sqrt{c(c+1)}-(2c+1-2\sqrt{c(c+1)})\SN^2(\frac{\sqrt{2c+1+2\sqrt{c(c+1)}}}{2}(\xi-\xi_1),
m_2) } {\CN^2(\frac{\sqrt{2c+1+2\sqrt{c(c+1)}}}{2}(\xi-\xi_1),
m_2)}}}.
 \end{eqnarray}\\

 According to the method and the above results , it is easy to prove that  all
 solutions  obtained in Ref.\cite{fu} can be represented by our solutions. For shortly, we
 only consider
 the solution $u_1$ in Ref.\cite{fu}, which has the
 following form
 \begin{equation}
u_1=2\sinh^{-1}(\pm\frac{m}{\sqrt{1-m^2}}\CN\xi),
 \end{equation}
 where $0<m<1$, and $\frac{\alpha}{k^2c}=-\frac{\alpha}{k\omega}=1-m^2.$
 Other cases can be proven similarly,  we omit them here.

 In fact, if we take traveling wave transformation $\xi=k(x-ct)$,
 then $\omega=-kc$. Under the first transformation of variable in
 Ref.\cite{fu}, the corresponding ordinary equation is given by
 \begin{equation}
(1+v^2)v''-v(v')^2+\alpha_1 v(1+v^2)^2=0,
 \end{equation}
 where $\alpha_1=\frac{\alpha}{k^2 c}$. Its general solution  is
 \begin{equation}
\pm(\xi-\xi_0)=\int
\frac{\mathrm{d}v}{\sqrt{-\alpha_1(v^2+1)(v^2-c_2)}},
 \end{equation}
 where $c_2$ is an integral constant. We take
 \begin{equation}
v=\frac{w-1}{2\sqrt w}, \ \ c_1=-(1+2c_2),
 \end{equation}
then corresponding equation is
\begin{equation}
\pm(\xi-\xi_1)=\int \frac{\mathrm{d}w}{\sqrt{-\alpha_1
w(w^2+2c_1w+1)}}.
 \end{equation}
We only consider the case $\alpha_1>0$, and the case $\alpha_1<0$
can be dealt with similarly. Then we have $c_2>v^2>0$, and hence
$c_1<-1$. From $-\sqrt {c_2}<v<\sqrt{c_2}$, we have
\begin{equation}
c_1-\sqrt{c_1^2-1}<-w<c_1+\sqrt{c_1^2-1}.
\end{equation}
Therefore the solution is
\begin{equation}
w=-c_1+\sqrt{c_1^2-1}-2\sqrt{c_1^2-1}
\SN^2(\frac{\sqrt{\alpha_1(\sqrt{c_1^2-1}-c_1)}}{2}(\xi-\xi_0),m_1),
\end{equation}
where $m_1^2=\frac{2\sqrt{c_1^2-1}}{\sqrt{c_162-1}-c_1}$. In
Ref.\cite{fu}, when $\alpha_1=1-m^2$ and $c_2=\frac{m^2}{1-m^2}$, Fu
\emph{et al} give solution
\begin{equation}
v_1=\pm\frac{m}{\sqrt{1-m^2}}\CN (\xi,m).
\end{equation}
In order to remove '$\pm$', we take $\xi_1$ such that $\CN
(\xi_1,m)=\pm 1$, and hence the above solution $v_1$ can be
rewritten as
\begin{equation}
v_1=\frac{m}{\sqrt{1-m^2}}\CN (\xi-\xi_1,m).
\end{equation}
Correspondingly, we have $c_1=-\frac{1+m^2}{1-m^2}$. From
$w_0=w(\xi_0)=\frac{(1+m)^2}{1-m^2}$, it follows that
$v(\xi_0)=\frac{w_0-1}{2\sqrt w_0}=\frac{m}{\sqrt{1-m^2}}$. On the
other hand, we have
$v(\xi_0)=\frac{m}{\sqrt{1-m^2}}\CN(\xi_0-\xi_1,m)$, so
$\CN(\xi_0-\xi_1,m)=1$. Under this condition, according to
$w=(v+\sqrt{1+v^2})^2$, we obtain a new identity of elliptic
functions
\begin{eqnarray}
\frac{(1+m)^2}{1-m^2}-\frac{4m}{1-m^2}\SN^2(\frac{1+m}{2}(\xi-\xi_0),\frac{4m}{(1+m)^2})\cr
=\{\frac{m}{\sqrt{1-m^2}}\CN
(\xi-\xi_1,m)+\sqrt{1+\frac{m^2}{1-m^2}\CN^2 (\xi-\xi_1,m)}\}^2.
\end{eqnarray}
This formula also give a new representation of solutions to
Sinh-G{\"o}rdon equation, i.e.,
\begin{equation}
u_1=\ln w=\ln (v+\sqrt{1+v^2})^2=2\sinh^{-1}v.
\end{equation}
According to the above method, other 36 solutions in Ref.\cite{fu}
can be represented by our solutions and furthermore obtain
corresponding new identities on Jacobian elliptic functions. For
shortly, we don't write those formulae concretely. In final, I would
like to point out that, in Ref.[38-40], by other transformations
like Landen's quadratic transformation, some identities on Jacobian
functions have been given.  But Our results are novel.

\section{Conclusions}
We obtain easily all single traveling wave atom solutions to
Sinh-G{\"o}rdon equation using direct integral method. Qualitative
properties of solutions are discussed, and some intuitive
conclusions are corrected. In order to prove that all other forms of
solutions in Ref.\cite{fu} can be represented by our solutions, we
discuss the solutions to elliptic equation. As a result, we can give
some new identities on Jacobian elliptic functions and show that all
solutions to Sinh-G{\"o}rdon equation in Ref.\cite{fu} can be
derived from our
solutions. By some examples, we display our method and results.\\

\textbf{Acknowledgements}: I would like to thank referee for his (or
her) suggestion for discussing qualitative properties of solutions
to Sinh-Gordon equation so that I can clarify those contradictories.

\end{document}